\documentclass[useAMS,usenatbib]{mn2e}
\usepackage{graphics,graphicx}
\usepackage{aas_macros}
\usepackage{amssymb}
\usepackage{amsmath} 
\usepackage{latexsym,hyperref}
\usepackage{color}
\usepackage{epstopdf}
\usepackage{marvosym}

\begin{document}

\title[The effect of water and electron collisions in the rotational
  excitation of HF in comets]{The effect of water and electron collisions in the rotational
  excitation of HF in comets }

\author[J. Loreau , A. Faure, and F. Lique]
{J.  Loreau$^{1}$ \thanks{E-mail: jerome.loreau@kuleuven.be}, 
A. Faure$^{2}$ \thanks{E-mail:alexandre.faure@univ-grenoble-alpes.fr}, 
F. Lique$^{3}$ \thanks{E-mail: francois.lique@univ-rennes1.fr}
  \\ $^1$ KU Leuven, Department of Chemistry, 3001 Leuven, Belgium
  \\ $^2$ Univ. Grenoble Alpes, CNRS, IPAG, F-38000 Grenoble, France
  \\ $^3$ Universit\'e de Rennes 1, F-35042 Rennes, France
}
  
\date{\today}

\pagerange{\pageref{firstpage}--\pageref{lastpage}} \pubyear{2022}

\maketitle

\label{firstpage}

\begin{abstract}
We present the first set of rate coefficients for the rotational excitation of the 7 lowest levels of hydrogen fluoride (HF) induced by collision with water molecules, the dominant collider in cometary comas, in the 5-150 K temperature range. The calculations are performed with a quantum statistical approach from an accurate rigid rotor \textit{ab initio} interaction potential. Rate coefficients for excitation of HF by electron-impact are also computed, within the Born approximation, in the $10-10,000$ K temperature range. These rate coefficients are then used in a simplified non-local thermodynamic equilibrium (non-LTE) model of a cometary coma that also includes solar radiative pumping and radiative decay. We investigate the range of H$_2$O densities that lead to non-LTE populations of the rotational levels of HF. 
We show that to describe the excitation of HF in comets, considering collisions with both water molecules and electrons is needed as a result of the large dipole of HF. 

\end{abstract}


\section{Introduction}

The interpretation of molecular spectra from astronomical sources is
often performed under the assumption of local thermodynamic
equilibrium (LTE). However, many astrophysical objects, including the
interstellar medium, protoplanetary disks and comets, display non-LTE
behaviour due to low particle densities ($n\ll 10^{10}$ cm$^{-3}$). An
accurate modelling of spectra then requires an understanding of the
balance between collision-induced transitions, spontaneous emission
and radiatively-induced transitions. Non-LTE radiative transfer models
rely on state-to-state rate coefficients for collisional energy
transfer that can be computed using a variety of theoretical
approaches. In the cometary coma, collisional excitation is generally
dominated by collisions with H$_2$O and electrons and, for comets at
large heliocentric distances, by CO \citep{BockeleeMorvan2004}. While
accurate rate coefficients for molecular excitation by collisions with
He or H$_2$ based on quantum-mechanical scattering calculations and 
\textit{ab initio} interaction potentials have been reported in the
literature, data are scarce for excitation induced by H$_2$O molecules. In that
case, the high density of rotational states of both colliders,
combined with a deep and anisotropic interaction potential, make fully
quantum-mechanical calculations challenging in terms of computational
time and memory requirements. State-to-state rate coefficients for
collisional excitation by water molecules are available for only a
handful of systems such as H$_2$O-CO \citep{Green1993, Faure2020},
H$_2$O-H$_2$O \citep{Buffa2000, Boursier2020}, and H$_2$O-HCN
\citep{Dubernet2019} and have all been obtained with an approximate
treatment of the dynamics.

In a recent work, we reported new rate coefficients for H$_2$O-CO
excitation and showed for a generic coma model that the non-LTE regime
occurred for H$_2$O densities in the range $10^3-10^8$ cm$^{-3}$
\citep{Faure2020}. Furthermore, we also showed that uncertainties in
the collisional data have significant influence on the CO population
distribution and that collisions with electrons have a negligible
impact except for unrealistically high electron fraction
\citep{Faure2022}. The present work focuses on the excitation of
hydrogen fluoride by water molecules and electrons. HF is thought to
be the main reservoir of fluorine in the interstellar medium
\citep{Neufeld2009}, and evidence suggests that in dense molecular
clouds, HF is present on the surface of grains rather than in the
gas-phase \citep{Phillips2010, Emprechtinger2012}. HF should thus be
present in comets formed from protostellar material. This has been
 tentatively confirmed by the far-infrared marginal detection of HF in the atmosphere of
comet C/2009 P1 Garradd with the \textit{Herschel} Space Observatory
\citep{BockeleeMorvan2014} and of comet 67P/Churyumov-Gerasimenko
through \textit{in situ} detection (via mass spectroscopy) with the
\textit{Rosetta} spacecraft \citep{Dhooghe2017}. The existence of
distributed sources of HF (and other halogen halides) in the form of
dust particles in the coma of 67P confirms that it is incorporated in dust grains
\citep{DeKeyser2017}. As HF is the best candidate to gain insights
into the abundance of halogens in comets, it is important to
investigate its collisional properties with the most abundant
colliders in comas, namely H$_2$O and free electrons.

In this work we present the first quantum state-to-state rate
coefficients for the rotational excitation of HF by H$_2$O. To that
end, we employ a recent \textit{ab initio} potential energy surface
combined with a quantum statistical approach. We also use the dipole
Born approximation to determine state-to-state rate coefficients for
the ro-vibrational excitation of HF by electrons. The computational
details are discussed in Section \ref{sec_theory} while the
HF$-$H$_2$O rate coefficients for excitation of the 7 lowest levels of
HF are discussed in Section \ref{sec_results}. Finally, in Section
\ref{sec_appli} we illustrate the impact of the new rate coefficients
by considering a non-LTE model of the excitation of HF in the cometary
coma that includes the effects of collision-induced transitions due to
both spin isomers of H$_2$O (ortho-H$_2$O and para-H$_2$O), electrons,
radiative pumping of the $v=0\to 1$ band due to solar photons, and
radiative decay due to spontaneous emission. We then discuss the relative effects of these sources of (de)excitation. We stress that our objective is not to revisit the HF abundance in comet C/2009 P1 Garradd, which is beyond the scope of this work, but only to evaluate the impact of collisional data on the non-LTE populations of HF.  Since HF has similarities with H$_2$O (mass and dipole moment), we also note that our results should be, at least partly, transferable to the excitation of H$_2$O in comets. Finally, our main conclusions are drawn in Section \ref{sec_concl}.

\section{HF collisional excitation rate coefficients}\label{sec_theory}

\subsection{HF$-$H$_2$O collisions}

The calculation of state-to-state rate coefficients for HF-H$_2$O
collisions relies on a recently-reported potential energy surface
(PES), on which the scattering is investigated.

\subsubsection{Potential energy surface}

We employed the high-level \textit{ab initio} PES recently constructed
by \citet{Loreau2020}. The HF and H$_2$O colliders were both treated
as rigid rotors, with geometrical structures corresponding to their
respective ground vibrational state. The interaction energy was
calculated on a five-dimensional grid (one distance and four angles)
by means of the explicitly-correlated single- and double-excitation
coupled cluster theory with a non-iterative perturbative treatment of
triple excitations [CCSD(T)-F12a] along with the augmented
correlation-consistent aug-cc-pVTZ basis sets.  The intermolecular
energies were subsequently fitted to a body-fixed expansion expressed
as a sum of products of spherical harmonics and a Monte-Carlo
estimator was employed to select only the largest expansion terms. This
includes in particular a description of the long-range, which is
required to compute collisional excitation rate coefficients at low
temperature.

The global minimum of the HF$-$H$_2$O PES occurs at a center-of-mass
separation of $R=4.99$ $a_0$ and corresponds to a non-planar
equilibrium structure in which a hydrogen bond is formed with the
water molecule acting as proton acceptor.  The depth of the PES at the
global minimum is $D_e = 3059$~cm$^{-1}$, while the dissociation
energy of the molecular complex is $D_0$= 2089.4 cm$^{-1}$ for
o-H$_2$O-HF and $D_0=2079.6$ cm$^{-1}$ for p-H$_2$O-HF.  Full details
about the potential surface calculations can be found in the paper by
\citet{Loreau2020}.  We note that recently a new HF$-$H$_2$O PES has
been presented by \citet{Viglaska2022} for non-rigid monomers and it
was found very similar to the PES used in the present work.

\subsubsection{HF$-$H$_2$O excitation rate coefficients}

The most accurate method to compute collisional excitation cross
sections and rate coefficients is the quantum-mechanical
close-coupling method. However, when the PES on which the collision
proceeds has a deep well and/or when the colliders have small rotational
constants, the large number of equations that need to be solved
simultaneously in order to reach convergence becomes prohibitive. In
the present case, we found it impossible to converge close-coupling
calculations even for a total angular momentum $J=0$, so that an
alternative approach was required.  We rely here on a statistical
approach of the dynamics based on a global \textit{ab initio} PES
\citep{Loreau2018a}. The validity of this method, inspired by the
statistical adiabatic model of \citet{Quack1975}, has been explored
for several systems including H$_2$O-CO \citep{Loreau2018c}. It is
expected to provide state-to-state rate coefficients that are correct
within a factor of 2 for collisions that proceed through a deep well
and at relatively low temperature, when the lifetime of the complex is
sufficiently long \citep{Konings2021, Balanca2020}.

The statistical method has been described for H$_2$O-CO collisions
elsewhere \citep{Faure2020} and is applicable to H$_2$O-HF, so that we
only recall its main features here.  The angular momenta of the two
colliding molecules (${\boldsymbol j_1}$ for H$_2$O and ${\boldsymbol
  j_2}$ for HF) are coupled to the angular momentum of the relative
motion ${\boldsymbol \ell}$ to form the total angular momentum
$\boldsymbol{ J = j_1+j_2+\ell}$.  The interaction Hamiltonian
excluding the kinetic term is then diagonalized in a basis of
rotational functions for the two colliding molecules. For each value
of the total angular momentum $J$, this leads to a set of adiabatic
rotational curves as a function of the intermolecular distance $R$
that can be connected asymptotically to the various rotational states of the
molecules. The collision is assumed to take place only if the
collision energy is larger than the height of the barrier in the
adiabatic curve corresponding to the initial state. The probability is
then divided equally among the channels for which the energy is larger
than their respective centrifugal barriers, which allows one to
compute state-to-state cross sections. These are then integrated over
a Maxwell-Boltzmann distribution of energies to obtain rate
coefficients.

The cross sections were calculated for transitions between all
rotational levels of both o-H$_2$O and p-H$_2$O up to $j_1=6$ and HF
up to $j_2=6$ and for total energies between 0 and 1500 cm$^{-1}$ in
order to calculate rate coefficients up to $T=150$ K. The basis set of
angular functions included levels up to $j_1=9$ and $j_2=10$, and
tests were performed to ensure the convergence of the results with
respect to the size of the basis set of rotational functions. The
rotational constant of HF corresponding to the ground vibrational level was taken to be $B_0=20.5567$ cm$^{-1}$ \citep{Webb1968},
while the energy $E_{j_1 k_a k_c}$ of the rotational levels of H$_2$O
was obtained using the effective Hamiltonian of \citet{Kyro1981}. They
are labeled with $j_1$, $k_a$, and $k_c$, where $k_a$ and $k_c$ are
pseudoquantum numbers corresponding to the projection of the angular
momentum $j_1$ along the axis of least and greatest moment of inertia,
respectively.  Convergence of the cross sections was obtained when
considering a maximum value of the total angular momentum
$J=150$. Adiabatic curves were generated for o-H$_2$O-HF and
p-H$_2$O-HF separately. Indeed, o-H$_2$O and p-H$_2$O colliders behave 
as two different species as they are not connected through inelastic collisions. 

Since, in the present work, we are only interested in the excitation of
HF, we summed the state-to-state rate coefficients over all possible
final states of H$_2$O and averaged over the initial rotational states
of H$_2$O assuming a thermal population distribution,
\begin{equation}\label{thermalized_rates}
k_{j_2\rightarrow j_2^\prime}(T)
= \sum_{j_1 k_a k_c j_1^\prime k_a^\prime k_c^\prime} P_{j_1 k_a k_c}(T) k_{j_1 k_a k_c, j_2 \rightarrow j_1^\prime k_a^\prime k_c^\prime , j_2^\prime}(T)
\end{equation}
\begin{equation}\label{eq_pop}
P_{j_1 k_a k_c}(T) = \frac{g_{j_1}\exp(-E_{j_1 k_a k_c}/k_BT)}{\sum_{j_1 k_a k_c} g_{j_1}\exp(-E_{j_1}/k_BT)}
\end{equation}
where $k_B$ is the Boltzmann constant and $g_{j_1}$ is the level
degeneracy. This set of ``thermalized'' rate coefficients for
(de-)excitation of HF assumes that the kinetic and H$_2$O rotational
temperatures are equal, and satisfies detailed balance. At the lowest
temperature considered here (10 K) the main contribution to the
thermalized rate coefficients originates from excitation by ground
state H$_2$O ($j_{k_a k_c}=0_{00}$ for p-H$_2$O and $1_{01}$ for
o-H$_2$O, respectively) while at higher temperatures the excited
rotational states of H$_2$O dominate.

\subsection{HF$-$electron collisions}

Scattering calculations for electron-HF collisions were performed
within the dipole Born approximation, which is briefly presented
below. We note that this approximation is employed in many models
treating excitation of molecules in comets \citep[e.g.][]{Zakharov2007}.

\subsubsection{The dipolar Born approximation}

Because HF has a strong dipole (1.83~D), dipole-allowed rotational
transitions ($\Delta j_2=\pm 1$) have very large cross sections, 
especially at low energy (below 1~eV). These cross sections are
dominated by high-partial waves, i.e. long-range interactions, so that
the dipole Born approximation is expected to be reliable.

The dipole Born integral cross section for a rotational excitation
$j_2\to j_2'$ can be formulated as follows \citep{Itikawa1972}:
\begin{equation}
  \sigma_{j_2\to j_2'}=\frac{8\pi}{3k^2}d^2\frac{S(j_2,
    j_2')}{2j_2+1}\ln\left|\frac{k+k'}{k-k'}\right|,
\end{equation}
where $k$ and $k'$ are the initial and final wave numbers of the
electron, respectively, $d$ is the permanent dipole moment and the
angular factor (or line strength) $S(j_2, j_2')$ is related to the
Einstein coefficient $A_{j_2' \to j_2}$ through:
\begin{equation}
A_{j_2'\to j_2}=\frac{64\pi^4\nu^3}{3hc^3}d^2\frac{S(j_2, j_2')}{2j_2'+1},
\end{equation}
where $h$ is the Planck's constant, $c$ is the speed of light and
$\nu$ is the frequency of the transition $j_2\to j_2'$. The Born
rotational cross section can thus be reformulated as:
\begin{equation}
  \sigma_{j_2\to
  j_2'}=\frac{hc^3}{8k^2\pi^3\nu^3}\frac{(2j_2'+1)}{(2j_2+1)}A_{j_2'\to
  j_2}\ln\left|\frac{k+k'}{k-k'}\right|.
\end{equation}
The equation is strictly similar for a ro-vibrational excitation
$v_2j_2\to v_2'j_2'$. In practice, Einstein coefficients were
extracted from the \texttt{HITRAN} database \citep{Gordon2017} for
dipole-allowed transitions among all levels below $v_2=2, j_2=0$ which
opens at 7750.8~cm$^{-1}$, i.e. the lowest 34 levels (up to $v_2=0,
j_2=19$).

For the rotational excitation of HF, \citet{Thummel1992} have
performed high-level calculations by using the variational $R$-matrix
method combined with the frame-transformation theory. They have shown
for the excitation $j_2=0\to 1$ that the Born rotational cross
sections are accurate to within a factor of 2 or better, as
expected from the large dipole. Good agreement between the Born approximation and
experimental rotational cross section for $j_2=0\to 1$ was also
emphasized by \citet{Itikawa2017}.

The vibrational excitation of HF was also studied experimentally and
theoretically \citep{Itikawa2017}. Strong threshold resonances in the
$v_2=0\to 1$ cross section were soon attributed to short-lived
negative-ion states \citep{Rohr1976}. Indeed, the dipole Born
vibrational cross section was found to be about an order of magnitude
smaller than the measured cross section (a few \AA$^2$ at threshold),
suggesting that the process is dominated by low-partial waves. We also
found a small $v_2=0\to 1$ Born cross section with a peak of $\sim$
0.1~\AA$^2$ around 1~eV. As a result, our Born cross sections were
multiplied by a factor of 10 so that the recommended $v_2=0\to 1$
cross sections of \citet{Itikawa2017} were reproduced to within a factor
of 2 beyond the threshold peak.

\subsubsection{HF-electron ro-vibrational rate coefficients}

Born cross sections were computed for electron energies up to 9~eV and
for all 59 dipole-allowed transitions $v_2j_2\to v_2'j_2'$. Rate
coefficients were deduced by integrating the cross sections over Maxwell-Boltzmann distribution of energies for kinetic temperatures between 10 and
10,000~K. For pure
rotational excitation, our rate coefficients should be accurate to
better than a factor of 2 while for ro-vibrational excitation their
accuracy is expected to be a factor of a few. Better accuracies
require higher-level treatments, including short-range effects, such
as R-matrix based calculations \citep[see e.g.][]{Ayouz2021}. We note,
however, that electron-impact vibrational excitation play a negligible
role in the application to comets presented below.


\section{Results and comparison of collision data sets}\label{sec_results}

\begin{figure}
\includegraphics*[width=8.5cm]{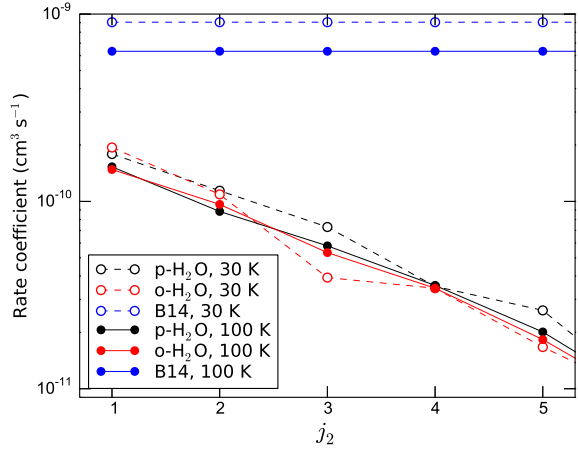}
\caption{ Rate coefficients for the rotational de-excitation of HF($j_2=1-5\rightarrow j_2^\prime=0$) by
  H$_2$O. The projectiles p-H$_2$O and o-H$_2$O are assumed to be thermalized at
  the kinetic temperature (30 or 100 K). The data obtained in the
  present work is compared to that used by \citet{BockeleeMorvan2014},
  denoted as B14.}
\label{fig_compB14_j0}
\end{figure}

Rate coefficients for the rotational de-excitation of HF($j_2=1-5\rightarrow j_2^\prime=0$) by
  H$_2$O are illustrated in Fig. \ref{fig_compB14_j0} for temperatures
of 30 K and 100 K. The dominant de-excitation rates occur for
$j_2=1$ with values larger than $10^{-10}$ cm$^3$ s$^{-1}$, and
the de-excitation rate coefficients decrease as $\Delta j_2$ increases. At $T=30$ K the large spacing
between energy levels of HF lead to rate coefficients that decrease
very quickly as $j_2$ increases. This observation remains true at $T=100$ K,
although the decrease with increasing $j_2$ is slightly less marked. The
rotational excitation of HF by H$_2$O depends on two counteracting
factors: the PES is deep and anisotropic, which should lead to large
excitation rate coefficients; however this is compensated by the large
rotational constant of HF.
At both temperatures we observe very little differences between the
excitation by p-H$_2$O and o-H$_2$O, which suggests that the
ortho-to-para ratio (OPR) of H$_2$O should not play a major role in
the excitation of HF, similar to what we reported for the excitation
of CO by H$_2$O \citep{Faure2020}.

\begin{figure}
\includegraphics*[width=8.5cm]{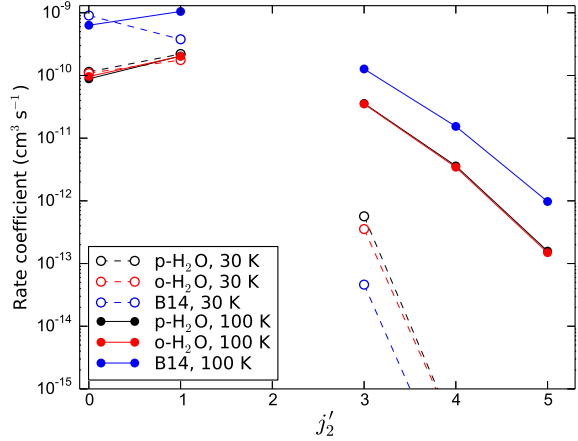}
\caption{Rate coefficients for the rotational excitation and de-excitation of HF($j_2=2\rightarrow j_2^\prime$).}
\label{fig_compB14_j2}
\end{figure}

For comparison, we display in Fig. \ref{fig_compB14_j2} the rate
coefficients for (de-)excitation from the level $j_2=2$. At both 30 K
and 100 K the dominant transition is the de-excitation towards
$j_2^\prime=1$, followed by the $j_2=2\rightarrow 0$ transition, while
the excitation rate coefficients quickly decrease with increasing $j_2^\prime$ due to large rotational spacing.

In Figs. \ref{fig_compB14_j0} and \ref{fig_compB14_j2}, we also compare
our results to the data used by \citet{BockeleeMorvan2014} (hereafter
denoted as B14) for the excitation of various molecules, including HF,
by H$_2$O. The B14 rate coefficients were obtained assuming a total
cross section $\sigma=5\times 10^{-14}$ cm$^2$, a value derived from
line broadening measurements on H$_2$O-H$_2$O. The total collisional
rate coefficient was computed as $k(T) = \sigma \langle v(T)\rangle$,
where $T$ is the kinetic temperature and $\langle v(T)\rangle$ is the
thermal mean velocity of the HF targets relative to the H$_2$O
projectiles. State-to-state rate coefficients $k_{j_2\rightarrow
  j_2^\prime}(T)$ were then obtained as $k_{j_2\rightarrow
  j_2^\prime}(T) = k(T) P_{j_2^\prime} (T )$, where $P_{j_2^\prime} (T
)$ is the Boltzmann population of the final level $j_2^\prime$ at
temperature $T$ \citep{BockeleeMorvan1987}. This is equivalent to
assuming that collisions redistribute the molecule according to the
Boltzmann distribution at a temperature given by the kinetic
temperature.

For both temperatures presented in Fig. \ref{fig_compB14_j0} the B14
rate coefficients are much larger than those obtained in the present work.
For de-excitation transitions to a given rotational level such as $j_2 \rightarrow j_2^\prime =0$, the difference between the two sets of rate coefficients increases with $j_2$ since the B14 data depend only on the final level $j_2^\prime$ but not on the initial level $j_2$, while the present results suggest a strong decrease with increasing $\Delta j_2$. For transitions $j_2 \rightarrow j_2^\prime =0$, this leads to discrepancies of a factor 5 (resp. 4) for $j_2=1$ at $T=30$ K (resp. 100 K) which increases with $j_2$ to reach a factor 26 (resp. 18) for $j_2=4$ at $T=30$ K (resp. 100 K).  
For transitions $j_2=2\rightarrow j_2^\prime$ at $T=100$ K the trends in
$j_2^\prime$ agree between the two sets of results, with the B14 rates
coefficients being larger by a factor 3-7 depending on $j_2^\prime$. 
On the other hand, at $T=30$ K we observe large discrepancies as the B14 rate coefficients are larger than those
computed here for HF de-excitation transitions, but smaller for transitions leading to excitation of HF. In addition, at low temperatures (e.g., 30 K)  our results predict a larger rate coefficient for the HF($2 \rightarrow 1$) transition than for HF($2 \rightarrow 0$), while the reverse is predicted by the B14 data since those rate coefficients only depend on the thermal population of the final state, which is larger for $j_2=0$ than $j_2=1$ at 30 K.

\begin{figure}
\includegraphics*[width=8.5cm]{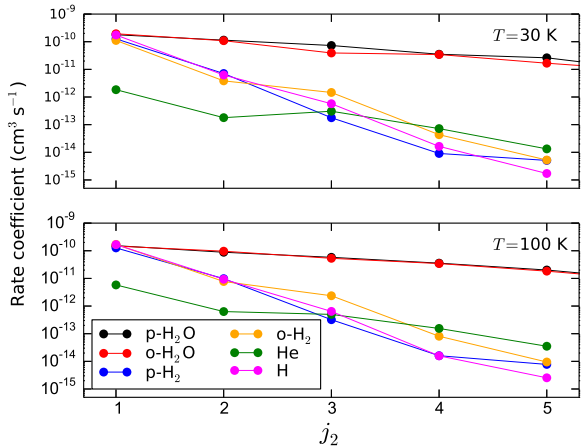}
\caption{ Rate coefficients for the rotational de-excitation of HF($j_2=1-5\rightarrow j_2^\prime=0$) by
  H$_2$O at two temperatures (30 K and 100 K), compared to the rate
  coefficients for de-excitation by collisions with H, He, p-H$_2$, and
  o-H$_2$.}
\label{fig_colliders}
\end{figure}

It is also instructive to compare the (de-)excitation of HF by other
colliders. Data is available for the (de-)excitation by p-H$_2$ and o-H$_2$
\citep{Guillon2012}, He \citep{Reese2005}, and H
\citep{Desrousseaux2018}. These data were all obtained with the fully
quantum-mechanical close-coupling method based on accurate \textit{ab
  initio} PESs. The comparison for transitions $j_2 \rightarrow
j_2^\prime =0$ is shown in Fig. \ref{fig_colliders} at two temperatures,
30 K and 100 K. With the exception of He, the rate coefficients
corresponding to the de-excitation $j_2=1\rightarrow 0$ are of similar
magnitude for all colliders. At 30 K the largest difference occurs
for de-excitation by o-H$_2$, with a rate coefficient that is a factor
$\sim 2$ smaller than for de-excitation by H$_2$O. At 100 K the
$j_2=1\rightarrow 0$ rate coefficients for the various colliders
besides He agree within 20\%, the largest difference occurring now
for the de-excitation by p-H$_2$.  On the other hand, for the (de-)excitation
towards/from more excited rotational levels ($\Delta j_2 > 1$) we observe
that H$_2$O is a much more efficient collider, with differences of up
to two orders of magnitude for the $j_2=3\rightarrow 0$
transition. This slower decrease of the rate coefficients with
increasing $\Delta j_2$ is a direct consequence of the deeper well and
large anisotropy of the HF$-$H$_2$O PES compared to the other systems
(the well depth $D_e$ of the H$_2$O-HF complex is 3059 cm$^{-1}$,
while it is 359 cm$^{-1}$ for HF-H$_2$, 135 cm$^{-1}$ for HF-H$_2$,
and 43.7 cm$^{-1}$ for HF-He).

\begin{figure}
\includegraphics*[width=8cm]{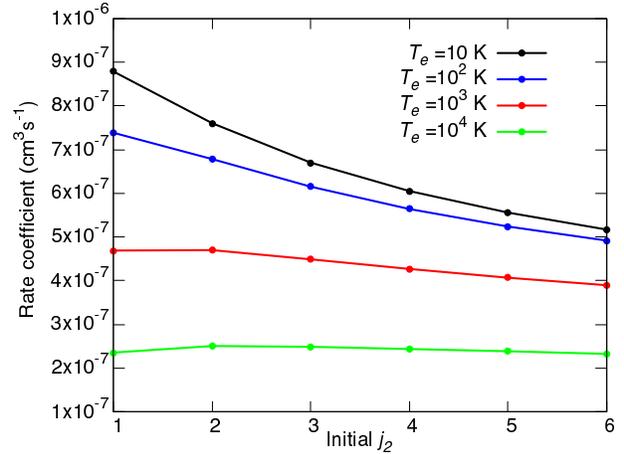}
\caption{Rate coefficients for the rotational de-excitation $j_2\to j_2-1$ of HF by
 electrons at four temperatures (10 K, 100 K, 1000 K, and 10000 K).}
\label{fig_exc_elec}
\end{figure}

The rate coefficients for de-excitation by electron impact are shown in Fig. \ref{fig_exc_elec} for dipolar transitions $j_2 \rightarrow j_2 -1$ for temperatures between 10 K and $10^4$ K. The large dipole moment of HF leads to rate coefficients that are in excess of  $10^{-7}$ cm$^3$s$^{-1}$ for all temperatures. At the lowest temperature (10 K) we observe a strong dependence of the rate coefficients on the initial value of $j_2$. This dependence decreases with increasing temperature to become almost negligible at $T=10^4$ K.

All de-excitation rate coefficients are available as supplementary
material and will be made available, along with the spectroscopic
data, in the \texttt{EMAA} database at
\href{http://emaa.osug.fr}{http://emaa.osug.fr} and in the \texttt{LAMDA} database at \href{https://home.strw.leidenuniv.nl/~moldata/}{https://home.strw.leidenuniv.nl/~moldata/}. Excitation rate coefficients can be
derived using the detailed balance principle.

\section{Non-LTE model of HF in comets}\label{sec_appli}

To assess the impact of the new collisional rate coefficients, we use
here the same non-LTE model of a cometary coma as in
\citet{Faure2020}. We compute the rotational level population of HF
based on a generic model of the coma that is not representative of a
particular comet. The comet is assumed to be located at a distance of
1~au from the Sun, with a total production rate
$Q_{\textrm{H$_2$O}}=10^{29}$~s$^{-1}$, an expansion velocity $v_{\textrm{exp}} = 0.8$~km
s$^{-1}$, and a uniform neutral gas temperature. Furthermore, we
assume that the collisional (de-)excitation of HF is due to both H$_2$O and
electrons, and that the water density profile follow a spherical Haser
model (as illustrated in Fig.~\ref{fig_temp_dens}).

For the electron density and temperature radial profiles, we use the
model described by \citet{Zakharov2007} where full details can be
found. The scaling factor $x_{ne}$ for the electron density is used as
a free parameter (see their Eq.~5). We note that the default value
$x_{ne} = 1$ was derived from Giotto measurements in comet
1P/Halley. Electron temperature and density profiles are shown in
Fig. \ref{fig_temp_dens}. As in the model used by
\citet{Zakharov2007}, the electron temperature is fixed to the gas
kinetic temperature for $R<R_{cs}$ and it increases linearly with $R$
up to $R=2R_{cs}$, where it reaches its maximum value of $T=10^4$
K. $R_{cs}$ is the distance of the `contact' surface (separating the
purely cometary plasma from the region containing solar wind charged
particles) and where $R_{rec}$ is the distance of the `recombination'
surface (outside which the recombination rate of electrons with ions
is negligible). We note that the bump in the electron density can be
partly attributed to the decrease of the recombination rate of
electrons with ions caused by the strong increase of the electron
temperature.

The coupled radiative transfer and statistical equilibrium (SE)
equations are solved with the public version of the \texttt{RADEX}
escape probability code, using the large velocity gradient (LVG)
approximation. In its public version, the electron and neutral
temperatures are identical. We have therefore implemented in
\texttt{RADEX} an electron temperature that can differ from the H$_2$O
kinetic temperature. The radiative transfer is treated as both local
and steady-state by running a grid of \texttt{RADEX} calculations
covering a range of H$_2$O densities (see below). We thus assume that
time-dependent effects are negligible and we also ignore HF
photodissociation (including photoionization). These two
approximations can be assessed by comparing the relative importance of
the different rates on the HF level population. As shown in
Fig.~\ref{fig_rates} for the HF level $j_2=1$, in the inner coma,
collisions with water molecules are largely dominant while in the
outer coma, the radiative processes (mainly the spontaneous emission
$j_2=1\to 0$) become the most important. In the intermediate region
between $10^2$~km and $10^4$~km, the contribution of electron
collisions is also significant. On the other hand, the
photodissociation and expansion rates have both a minor contribution
over the entire coma. As a result, both photochemical and
time-dependent effects should have a negligible impact on the HF
populations and the usual steady-state assumption of SE is expected to be
reliable (see details in Appendix~A).

The input parameters to \texttt{RADEX} are the kinetic temperature,
$T_k$, the column density of HF, $N$(HF), the line width, $\Delta v$,
and the density of the colliders, here $n_{\rm H_2O}$ and $n_{\rm
  e^-}$. A representative kinetic temperature of 50~K was chosen. The
column density was fixed at $N({\rm HF})=10^{11}$~cm$^{-2}$, which is
the typical magnitude at $r\sim 1000$~km for a HF/H$_2$O abundance
ratio of $\sim 1.8\times 10^{-4}$ \citep{BockeleeMorvan2014}. This
column density corresponds to line opacities lower than 0.1. The
background radiation field includes both the 2.73~K cosmic microwave
background (CMB) (of negligible importance) and the Sun
radiation. Other parameters are taken as in \citet{Faure2020}, where
details are provided.

Input data include the HF energy levels $(v_2, j_2)$, the spontaneous
emission Einstein coefficients and the collisional rate
coefficients. Level energies and Einstein coefficients were taken from
the \texttt{HITRAN} database, as for the electron-impact excitation
calculations described above. Only the first excited vibrational level
$v_2=1$ was taken into account. Our model thus include the lowest 34
ro-vibrational levels of HF, i.e. up to level $(0, 19)$ which lies
7515.3~cm$^{-1}$ above the ground-state $(0, 0)$. Since our H$_2$O-HF
SACM rate coefficients are available for the lowest 7 rotational
levels only, we had to extrapolate the H$_2$O-HF collisional data
above $j_2=6$. The method implemented by \citet{Faure2020} for
H$_2$O-CO was employed here for H$_2$O-HF. It should be noted,
however, that the corresponding levels are negligibly populated (less
than 1\%) over the entire coma so that the extrapolated collisional
data play a negligible role in the SE equations.

\begin{figure}
\includegraphics*[width=8.5cm]{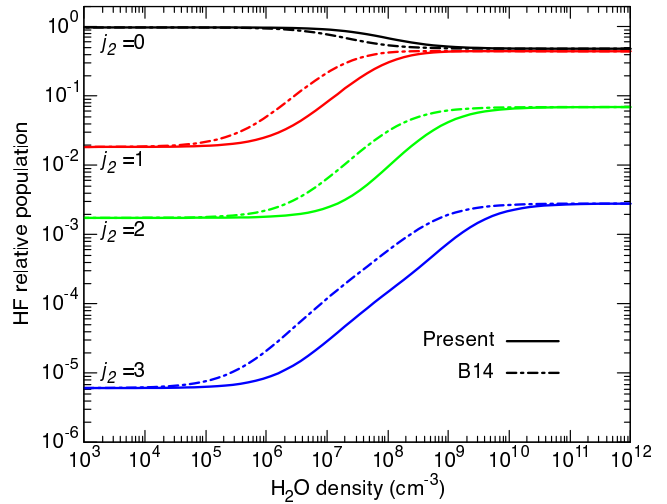}
\caption{Level populations of HF ($j=0-3$) as functions of H$_2$O
  density for our non-LTE model at a temperature of 50 K. Two sets of
  collisional data for H$_2$O collisions are employed: the thermalized
  SACM set with the OPR of H$_2$O fixed at 3 (solid lines) and the B14
  set (dashed lines). Electron collisions are neglected by setting
  $x_{n_e}=0$. See text for details.}
\label{fig_pop_noelec}
\end{figure}

In a first instance we consider only excitation by H$_2$O and neglect
the contribution of electrons by setting $x_{n_e}=0$. The population
of HF levels $j_2$ obtained with our non-LTE model based on the set of
thermalized SACM rate coefficient is plotted as a function of the
H$_2$O density in Fig.~\ref{fig_pop_noelec} for a kinetic temperature
of 50~K and with an OPR for H$_2$O set to 3.
We can clearly observe the presence of three different regimes: the
fluorescence equilibrium at low densities ($n_{\rm H_2O}\lesssim
10^5$~cm$^{-3}$), the non-LTE regime in the range $10^5\lesssim n_{\rm
  H_2O}\lesssim 10^{10}$~cm$^{-3}$ and the LTE-regime at larger
densities. Based on the isotropic Haser model and a production rate
$Q_{\rm H_2O}= 10^{29}$~s$^{-1}$, this implies that LTE applies for HF
at $r\lesssim 30$~km, i.e. in the inner coma, while the fluorescence
equilibrium distribution is completely established at $r\gtrsim 10^4$~km (see Fig.~\ref{fig_temp_dens}). The non-LTE regime thus
extends typically from 30 to 10,000~km. This can be compared to the
collisional excitation of CO, for which the non-LTE regime extends
from 300 to 70,000 km (corresponding to $10^3\lesssim n_{\rm
  H_2O}\lesssim 10^{8}$~cm$^{-3}$) for identical coma parameters
\citep{Faure2020}. This difference mainly reflects the larger
spontaneous emission rates in HF. We note that in the fluorescence
equilibrium regime, only the ground rotational state of HF is
significantly populated at a kinetic temperature of 50~K (the
population of $j_2=1$ is about 2\%), while at thermal equilibrium the
population of the $j_2=0-2$ levels is 48\%, 44\%, and 7\%,
respectively.

On Fig.~\ref{fig_pop_noelec} we also show the rotational population
obtained with the B14 set of rate coefficients.  Since the rate
coefficients computed in this work are smaller than those of
\citep{BockeleeMorvan2014}, one of the main differences between the
two sets of data is that the LTE equilibrium is predicted to be
reached at H$_2$O densities almost one order of magnitude higher than
if the B14 rate coefficients are employed. The population of ground
state HF ($j_2=0$) is also predicted to be higher than with the B14
set, while the population of excited rotational states is smaller by
up to a factor 3. A similar effect was observed for the excitation of
CO by H$_2$O \citep{Faure2020}.

\begin{figure}
\includegraphics*[width=8.5cm]{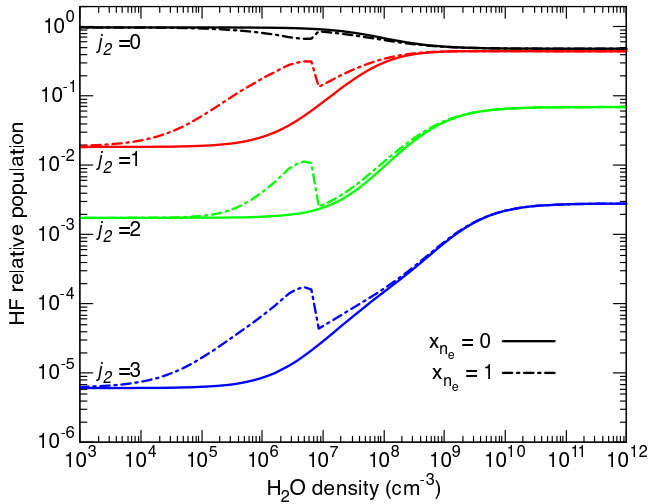}
\caption{Level populations of HF ($j=0-3$) as functions of H$_2$O
  density for our non-LTE model. The collisional data are the
  thermalized SACM rate coefficients for H$_2$O collisions with the
  OPR of H$_2$O fixed at 3 (solid lines) complemented with the Born
  rate coefficients for electron collisions (dot-dashed lines). The
  scaling factor $x_{ne}$ is taken equal to 1. See text for details.}
\label{fig_pop_withelec}
\end{figure}

We now consider the impact of excitation by electrons. The population
of the HF rotational levels including excitation by electron-impact is
shown in Fig.~\ref{fig_pop_withelec}. It can be seen that electrons
play a major role for HF at H$_2$O densities around $5\times 10^6$
cm$^{-3}$, leading to a decrease of the population of the ground state
$j_2=0$ an a strong increase of the population of excited rotational
states.  The largest impact occurs due to the jump in electron density
at a distance of about 2000~km from the comet nucleus (see
Fig.~\ref{fig_temp_dens}). We conclude that electron-impact excitation
play a significant role, even for a standard i.e. moderate value of
the scaling factor ($x_{n_e}=1$). The importance of electron
collisions for polar species was indeed emphasized in previous works
\citep{Xie1992}. This again supports
the need for higher-level electron-HF calculations in order to confirm, and
possibly improve, the present results.

Regarding the (marginal) detection of HF $j_2=1 \to 0$  in comet C/2009 P1 (Garradd) with Herschel \citep{BockeleeMorvan2014}, our results show that using the SACM HF-H$_2$O rate coefficients will require a larger HF abundance than using the B14 rate coefficients, whatever the excitation model. Indeed, as plotted in Figure~5, the population of $j_2=1$ is significantly reduced at intermediate densities when the present (smaller) rate coefficients are employed. Of course the actual impact of the new rate coefficients depends on the contribution of the intermediate density region to the overall flux within the Herschel beam. It thus depends on the exact physical model used for C/2009 P1 and, in particular, on the gas temperature profile which was taken as variable in \citet{BockeleeMorvan2014}. Such a detailed modeling is clearly beyond the scope of this work, but our results suggest that the HF/H$_2$O abundance ratio ($1.8 \pm 0.5 \times 10^{-4}$) derived by \citet{BockeleeMorvan2014} and used as a proxy for the F/O elemental ratio is a lower limit. Within the uncertainty margins, it may still be consistent with values derived for other Solar System objects (see Table~3 in \cite{Dhooghe2017}), in particular the F/O ratio measured with {\it Rosetta} in comet 67P which varies between $0.2-4\times 10^{-4}$, with a weighted average bulk value of $0.89\times 10^{-4}$. Clearly, only a future far-IR observatory in space or at airborne altitude will be able to confirm or refute the identification of HF in comet C/2009 P1. If confirmed, the present set of SACM collisional data will be essential to derive an accurate F/O ratio.

\section{Concluding remarks}\label{sec_concl}

In this work we obtained the first quantum state-to-state rate coefficients corresponding to the rotational excitation of HF by p-H$_2$O and o-H$_2$O for temperatures up to 150 K and rotational levels up to $j_2=6$. The calculations were performed by means of a quantum statistical approach based on a recent five-dimensional \textit{ab initio} potential energy surface obtained from highly correlated approaches. 

The state-to-state rate coefficients were summed over the final states of H$_2$O and averaged over the initial rotational states of H$_2$O assuming a rotational population corresponding to equilibrium at the kinetic temperature to provide thermalized rate coefficients. The comparison with approximate data calculated in previous works using simple assumptions showed large differences at all temperatures considered. 
We also found H$_2$O to be a more efficient collider than H$_2$, H, or He in the excitation of HF, and the excitation by o-H$_2$O or p-H$_2$O to be very similar.

The data were then used in a generic non-LTE model of the cometary coma. We showed that the rotational population of HF is strongly affected by collisional excitation for H$_2$O densities in the range between $10^5$~cm$^{-3}$ and $10^{10}$~cm$^{-3}$. 

Moreover, we provided rate coefficients for the excitation of HF by electron in the Born approximation. The corresponding rate coefficients were found to be large ($>10^{-7}$~cm$^3$s$^{-1}$) as a result of the large dipole moment of HF. This suggests that electrons will compete with neutrals in exciting HF when the electron-to-neutral number density ratio is larger than $10^{-3}$. We therefore considered the impact of excitation by electrons in our coma model, and showed that this process plays an important role in the rotational population of HF, with the largest impact observed at a distance of about 2000~km from the comet nucleus, a result of the jump in electron density.

The present results suggest that a non-LTE modelling of the excitation of HF and other molecular species in the cometary coma is required over a wide range of distance to the nucleus, and that accurate excitation rate coefficients are needed. 

\section*{Acknowledgements}
J.L. acknowledges support from Internal Funds KU Leuven through Grant
No. 19-00313.  F.L. acknowledges Rennes Metropole for financial support. This research was supported by the CNRS national
program `Physique et Chimie du Milieu Interstellaire'. Nicolas Biver
is acknowledged for useful discussions.

\section*{Data availability}
The data underlying this article are available in the article and in its online supplementary material.

\appendix

\section{Validation of the time-independent model}

\begin{figure}
\includegraphics[width=8.5cm]{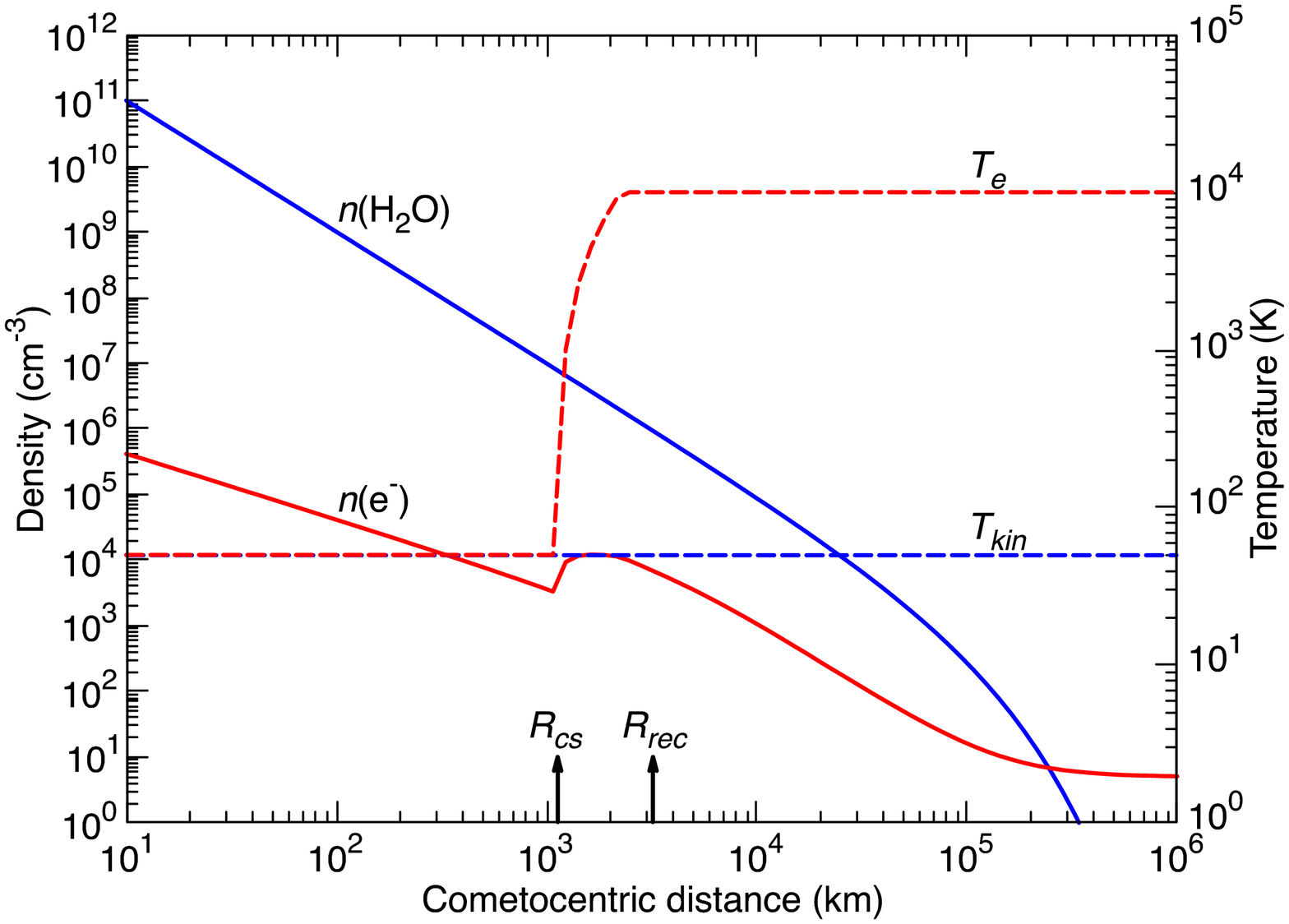}
\caption{Coma density (solid lines) and temperature (dashed lines)
  profiles of H$_2$O (blue) and electrons (red) used in our non-LTE
  model. See text for details.}
\label{fig_temp_dens}
\end{figure}

In our steady-state non-LTE model, the H$_2$O density profile follows
a simple Haser model, as shown in Fig.~\ref{fig_temp_dens}, with a
uniform gas kinetic temperature of 50~K. The electron and density
temperature profiles were taken from the model of \citet{Zakharov2007}
where full details can be found.

\begin{figure}
\includegraphics[width=8.5cm]{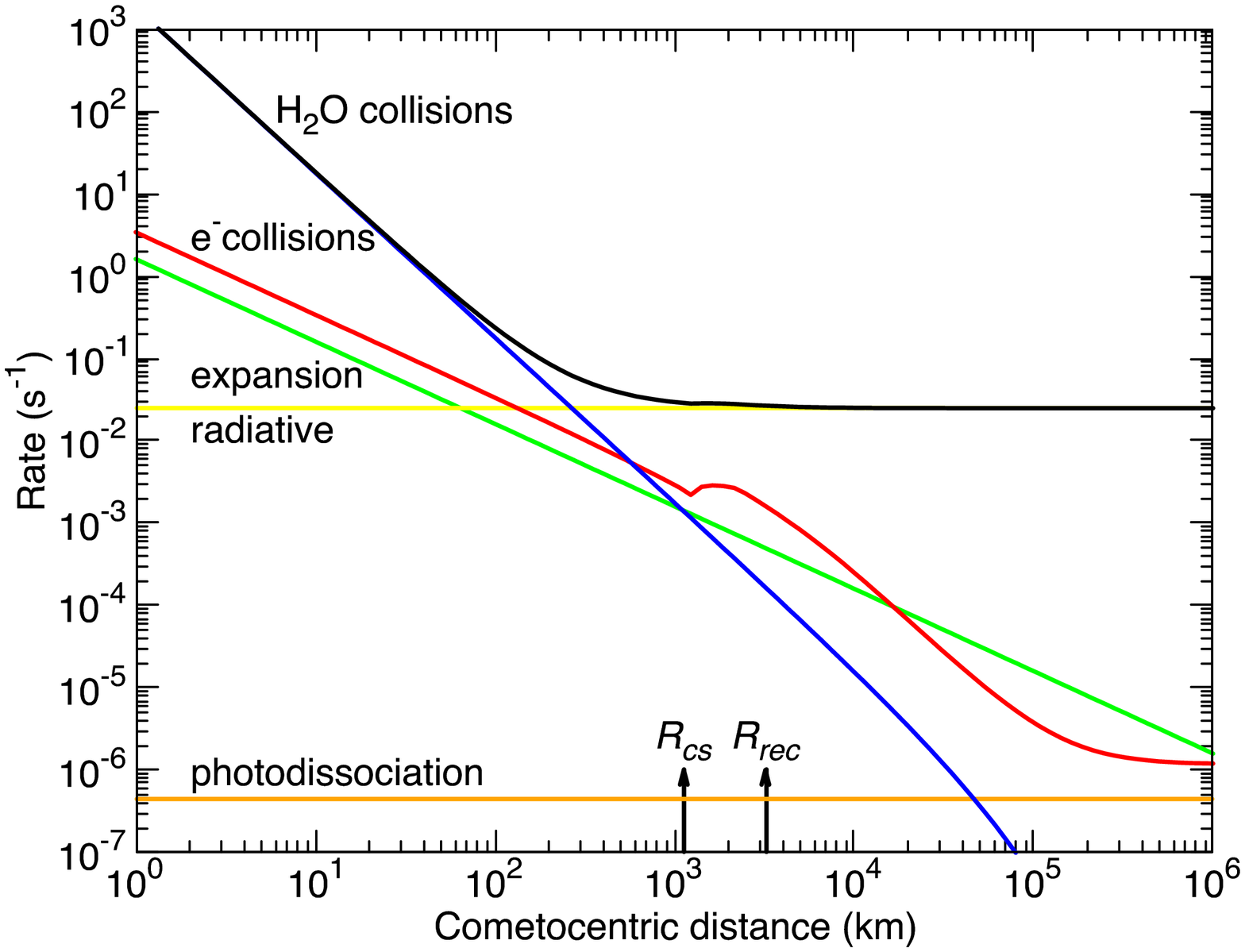}
\caption{The different rates (in s$^{-1}$) for the processes involved
  in the depopulation of the HF level $j=1$ as functions of the
  cometocentric distance. The black solid line denotes the sum of
  collisional and radiative rates.}
\label{fig_rates}
\end{figure}

In order to assess the validity of the steady-state assumption to
solve the SE equations, we plot in Fig.~\ref{fig_rates} the different
rates involved in the depopulation of the HF $j_2=1$ level:
collisional de-excitation by H$_2$O molecule and electrons (present
data), spontaneous emission and infrared pumping (data taken from
\texttt{HITRAN} database), coma expansion ($\gamma=2v_{exp}/R$ with
$v_{exp}$=0.8~km.s$^{-1}$) and photodissociation ($\beta=4.4\times
10^{-7}$~s$^{-1}$ from \citet{Huebner2015}). The black solid line
denotes the sum of collisional and radiative rates. Note that the
contribution of IR pumping is negligible. It can be observed that the
photodissociation and expansion processes both have minor
contributions at all cometocentric distances so that the standard steady-state
approximation of SE should be reliable over the entire coma.

It should be noted that similar results were recently observed by
\citet{Bergman2022} in the case of HCN. By running time-dependent
radiative transfer calculations, these authors were able to quantify
these effects and they have shown that the photodissociation and
expansion processes do actually modify the HCN level populations by
less than 15\%. Even smaller deviations are expected in the case of HF
since its spontaneous emission rates are much larger than those of
HCN. More generally, photochemical and time-dependent effects should
be moderate for molecules with sizeable dipole moments owing to their
large radiative rates and large electron-impact excitation rate
coefficients. The steady-state assumption may become questionable for
molecules with small dipoles (e.g. CO) or for comets with small water
production rates (or small electron density).

\label{lastpage}

\end{document}